# Gender Politics in the 2016 U.S. Presidential Election: A Computer Vision Approach

Yu Wang
Political Science
University of Rochester
Rochester, NY, United States

Yang Feng, Jiebo Luo*
Computer Science
University of Rochester
Rochester, NY, United States

Xiyang Zhang
Psychology
Beijing Normal University
Beijing, China

## ABSTRACT
Gender is playing an important role in the 2016 U.S. presidential election, especially with Hillary Clinton becoming the first female presidential nominee and Donald Trump being frequently accused of sexism. In this paper, we introduce computer vision to the study of gender politics and present an image-driven method that can measure the effects of gender in an accurate and timely manner. We first collect all the profile images of the candidates' Twitter followers. Then we train a convolutional neural network using images that contain gender labels. Lastly, we classify all the follower and unfollower images. Through two case studies, one on the 'woman card' controversy and one on Sanders followers, we demonstrate how gender is informing the 2016 presidential election. Our framework of analysis can be readily generalized to other case studies and elections.

## 1. INTRODUCTION
Gender has always played an important role in American elections (e.g. Ronald Reagan's re-election in 1984, George W. Bush's election in 2000 [21], Barack Obama's election in 2008 and re-election in 2012 [11]). It is set to play an important role again in the 2016 election cycle. On the Democrats' side, Hillary Clinton has become the first female presidential nominee for a major political party in the U.S. history. On the Republican side, Donald Trump is frequently accused of sexism, with his controversies against Carly Fiorina, Megyn Kelly, Heidi Cruz, and Hillary Clinton. Naturally, being able to measure the effects of gender in an accurate and timely manner becomes crucial.

*Jiebo Luo (jluo@cs.rochester.edu) is the corresponding author of this paper.

Recent advances in computer vision [14, 22, 23, 7] have made object detection and classification increasingly accurate. In particular, face detection and gender classification [4, 10, 16] have both achieved very high accuracy, largely thanks to the adoption of deep learning [15] and the availability of large datasets [9, 12, 20] and more recently [6].

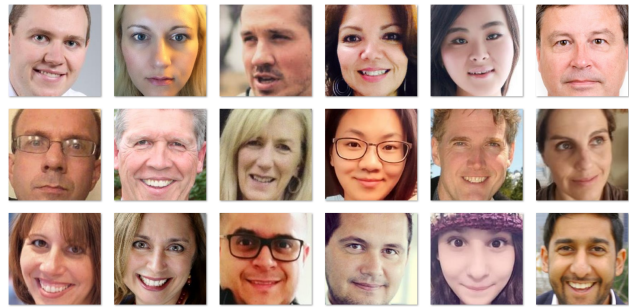

Figure 1: Top row: Hillary Clinton Unfollowers; Middle: Donald Trump Followers; Bottom: Bernie Sanders Followers.

In this paper, we introduce computer vision to the study of gender politics and present an image-driven method to analyze how gender is shaping the 2016 presidential election. We first collect the profile images of the candidates' followers and unfollowers (Figure 1).[1] Then we select the images with gender labels to train a convolutional neural network and we use the trained network to classify all the images for Hillary Clinton, Bernie Sanders, and Donald Trump. Lastly, we construct a gender affinity model and test statistically whether or not an event of interest has disturbed the prior gender balance.

To illustrate the effectiveness of our method, we select two case studies that carry great importance: (1) the effects of the 'woman card' controversy, and (2) the effects of gender on Bernie Sanders followers, who now need to decide whether to support Clinton or Trump. The 'woman card' controversy refers to the incident where Trump accused Hillary Clinton of playing the 'woman card' against him. The Bernie Sanders case refers to the reports that after Sanders lost the Democratic primary to Clinton, his male supporters have a higher probability of voting for Trump than females. We provide more background information in Section IV.

---
[1] By 'unfollower', we mean people who previously followed a candidate on Twitter and later unfollowed him/her.

In the first case, we show that the 'woman card' controversy has made women more likely to follow Hillary Clinton and less likely to unfollow her. In the second case, we demonstrate those Sanders followers who are likely to switch to Trump are predominantly male. Our framework of analysis, which marries gender politics with computer vision, can be readily generalized to study other cases and even other elections, such as the French presidential election in 2017.

## 2. RELATED LITERATURE

Our work builds on previous literature in electoral studies, data mining, and computer vision.

In eletoral studies, researchers have argued that gender constitutes an important factor in voting behavior. One common observation is that women tend to vote for women, which is usually referred to as gender affinity effect [13, 3, 1]. Another observation is that pre-election polls tend to underestimate support for female candidates [24]. In the 2016 presidential election, Hillary Clinton explicitly portrays herself as a champion "fighting for women's healthcare and paid family leave and equal pay." It is also widely reported that male Sanders supporters are more likely to vote for Trump than female Trump supporters. Our work will test the strength of this gender affinity effect by constructing a random utility model.

In data mining, there is a burgeoning literature on using social media data to analyze and predict elections. In particular, several studies have explored ways to infer users' preferences. According to [19], tweets with sentiment can potentially serve as votes and substitute traditional polling. [28] exploits the variations in the number of 'likes' of the tweets to infer Trump followers' topic preferences. [17] uses candidates' 'likes' in Facebook to quantify a campaign's success in engaging the public. [27] uses follower growth on the dates of public debate to measure candidates' debate performance. Our work also pays close attention to the number of followers, but we go further by investigating the gender composition of these followers.

Our work ties in with current computer vision research. In this dimension, our work is related to gender classification using facial features. [16] uses a five-layer network to classify both age and gender. [5] introduces a dataset of frontal-facing American high school yearbook photos and uses the extracted facial features to study historical trends in the U.S. [10] collects 4 million weakly labeled images to train an SVM classifier and has achieved an accuracy of 96.98%. [26] uses user profile images to study and compare the social demographics of Trump followers and Clinton followers. [25] focuses specifically on the unfollowers of Donald Trump and Hillary Clinton, and reports that women are more likely to unfollow both candidates.

## 3. DATA AND METHODOLOGY

### 3.1 Data

In this section, we describe our dataset *US2016*, the pre-processing procedures and our CNN model. One key variable is *number of followers*. This variable is available for all three candidates and covers the entire period from September 18, 2015 to Oct 19, 2016. Compared with other candidates who have dropped out of the race, the two presidential nominees (and Bernie Sanders) also have the most Twitter followers (Figure 2). This variable is updated every 10 minutes in our program. In Figures 2 and 3, we present the cumulative number of Trump followers and Clinton followers and the net follower gain respectively.[2] By recording the follower IDs in a timely manner, we are able to identify the new followers and the unfollowers:

#(net follower gain) = #(new followers) − #(unfollowers)

Besides the number of followers, our dataset *US2016* also contains the detailed follower IDs for Trump, Clinton and Sanders on specific dates, including March 24th, April 17th and May 10th. This information enables us to track the evolution of the election dynamics.

### 3.2 Modeling Gender Affinity

We hypothesize that gender-related events have asymmetrical effects on men and women. When such an event occurs, it will disturb the gender balance previously observed. In the context of Twitter following, this can be modeled formally as:

$$U_m = \beta X_m + \lambda_m E + \epsilon$$

$$U_w = \beta X_w + \lambda_w E + \epsilon$$

where X is a vector of static variables related to one's following inclination, such as education, age, and income, $\lambda_m$ represents the impact of the event $E$ on a man, $E$ denotes the occurrence of an event and is binary, $\lambda_w$ is the utility impact on a woman, $\epsilon \sim Normal(1,0)$, and $U$ denotes the utility of following. Individuals will follow a candidate if and only if their utility of following is positive.

This translates into a probability of following for men and women respectively as follows:

$$Pr(Y_m = 1) = \Phi(\beta X_m + \lambda_m E)$$

$$Pr(Y_w = 1) = \Phi(\beta X_w + \lambda_w E)$$

Therefore, the gender distribution of new followers prior to an event is calculated as:

$$\frac{N'_m \Phi(\beta X_m)}{N'_w \Phi(\beta X_w)}$$

where $N'_m$ is the number of prospective male followers and $N'_w$ is the number of prospective female followers for the period before the event. After the event has occurred, the gender distribution of new followers becomes:

$$\frac{N''_m \Phi(\beta X_m + \lambda_m)}{N''_w \Phi(\beta X_w + \lambda_w)}$$

where $N''_m$ is the number of prospective male followers and $N''_w$ is the number of prospective female followers in the period immediately after the event.

---
[2]For a detailed analysis of follower growth patterns, see [27].

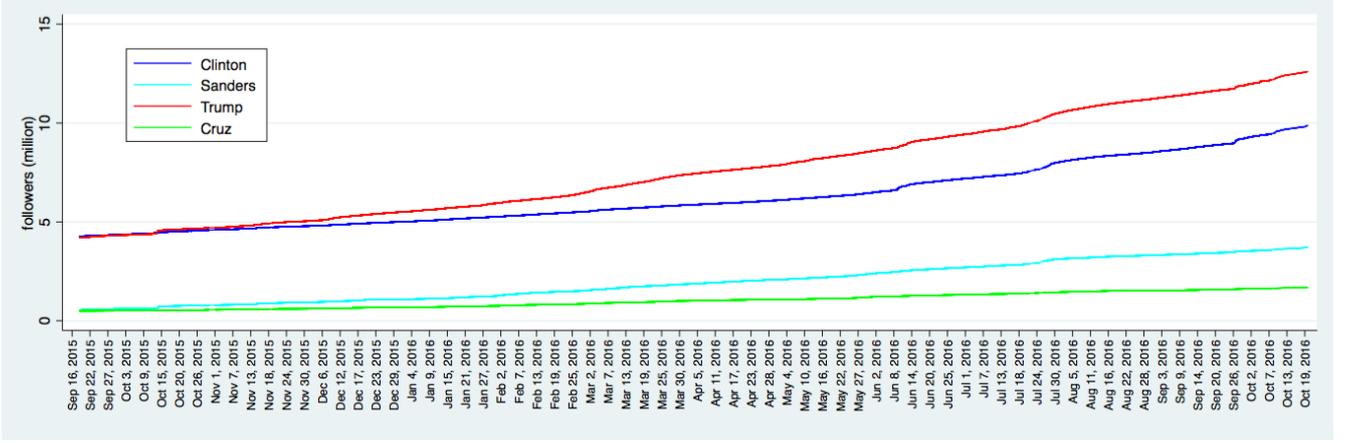

Figure 2: Cumulative follower growth for Hillary Clinton, Bernie Sanders, Donald Trump and Ted Cruz.

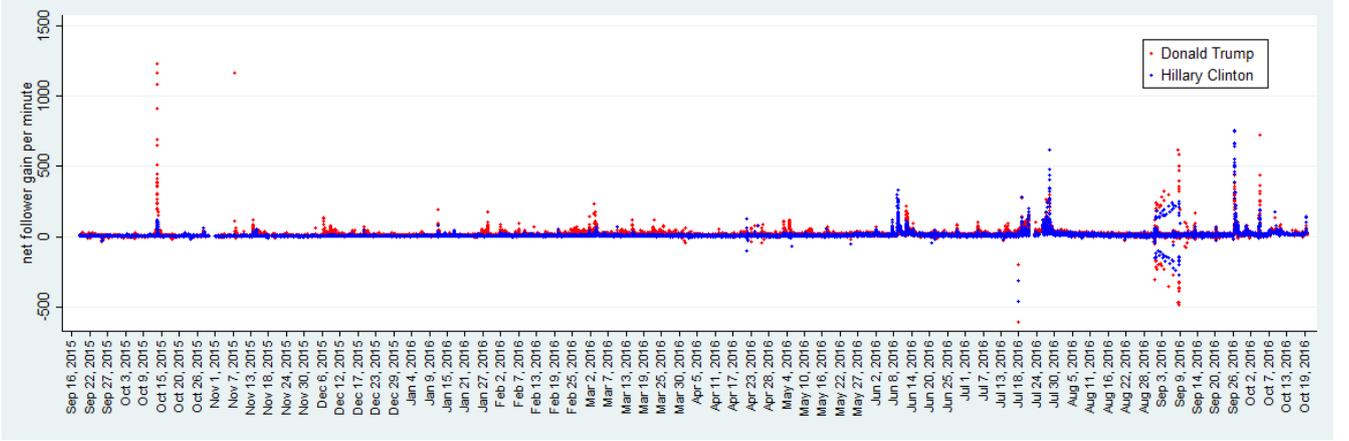

Figure 3: Net follower gain per minute for Donald Trump and Hillary Clinton.

Finally, the disturbance in the gender distribution that is attributed to the event is calculated as:

$$D(Event) = \frac{N''_m}{N''_w} \times \frac{\Phi(\beta X_m + \lambda_m)}{\Phi(\beta X_w + \lambda_w)} - \frac{N'_m}{N'_w} \times \frac{\Phi(\beta X_m)}{\Phi(\beta X_w)}$$

### 3.3 Statistical Testing

For an event that could disproportionately affects women, such as the 'woman card' controversy, we expect that there will be a positive disturbance towards men in the gender distribution among Trump followers and that the disturbance will tilts towards women for Hillary Clinton. The statistical significance of disturbance $D(Event)$ can then be calculated using the two-sample z-test:

$$z = \frac{\hat{p}_2 - \hat{p}_1}{\sqrt{\hat{p}(1-\hat{p})(1/n_2 + 1/n_1)}}$$

where

$$\hat{p}_2 = \frac{N''_m \times \Phi(\beta X_m + \lambda_m)}{N''_m \times \Phi(\beta X_m + \lambda_m) + N''_w \times \Phi(\beta X_w + \lambda_w)}$$

$$\hat{p}_1 = \frac{N'_m \times \Phi(\beta X_m + \lambda_m)}{N'_m \times \Phi(\beta X_m) + N'_w \times \Phi(\beta X_w)}$$

$$n_2 = N''_m \times \Phi(\beta X_m + \lambda_m) + N''_w \times \Phi(\beta X_w + \lambda_w)$$

$$n_1 = N'_m \times \Phi(\beta X_m) + N'_w \times \Phi(\beta X_w)$$

$$\hat{p} = \frac{n_1 * \hat{p}_1 + n_2 * \hat{p}_2}{n_1 + n_2}$$

With large $n_1$ and $n_2$, by the central limit theorem z is approximately standard normal. The null hypothesis is that the event of interest has not disturbed the gender balance among Twitter followers and unfollowers. A large z in absolute terms (2.1 for example) will be strong evidence that there exists a disturbance and that the null hypothesis should be rejected.

### 3.4 Gender Inference by Computer Vision

Furthermore, we collect the profile images based on follower IDs. Our goal is to infer an individual's gender based on the profile image and to test the hypothesis that gender is affecting the following and unfollowing behavior of the presidential candidates' Twitter followers.

To process the profile images, we first use OpenCV to identify faces, as the majority of profile images only contain a face.[3] We discard images that do not contain a face and the ones in which OpenCV is not able to detect a face. When multiple faces are available, we choose the largest one. Out of all facial images thus obtained, we select only the large ones. Here we set the threshold to 18kb. This ensures high image quality and also helps remove empty faces. Lastly we resize those images to (28, 28, 3).

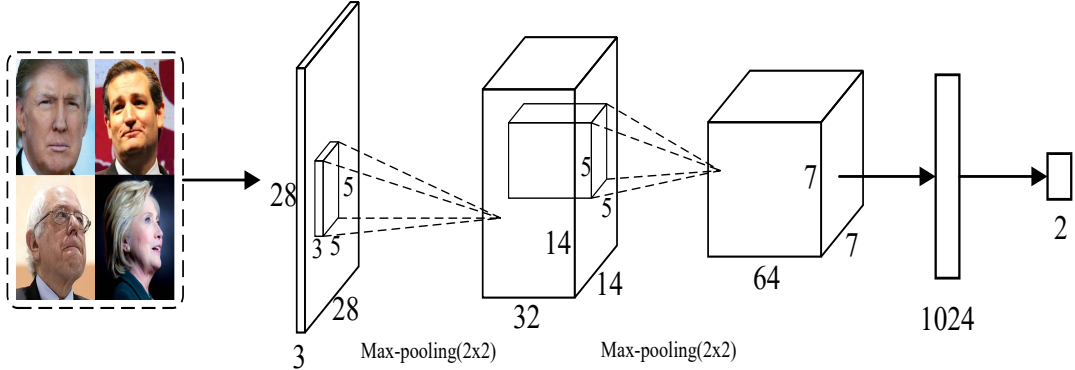

Figure 4: The CNN model consists of 2 convolution layers, 2 max-pool layers and a fully connected layer.

To classify profile images, we train a convolutional neural network using 42,554 weakly labeled images, with a gender ratio of 1:1. These images come from Trump's and Clinton's current followers. We infer their labels using the followers' names. For example, David, John, Luke and Michael are male names, and Caroline, Elizabeth, Emily, Isabella and Maria are female names.[4] For validation, we use a manually labeled data set of 1,965 profile images for gender classification. The validation images come from Twitter as well so that we can avoid the cross-domain problem. Moreover, they do not intersect with the training samples as they come exclusively from individuals who unfollowed Hillary Clinton before March 2016.

Table 1: Summary Statistics of CNN Performance

| Architecture | Precision | Recall | F1 | Accuracy |
|---|---|---|---|---|
| 2CONV-1FC | 91.36 | 90.05 | 90.70 | 90.18 |

The architecture of our convolutional neural network is illustrated in Figure 4, and we are able to achieve an accuracy of 90.2%, which is adequate for our task ( Table 1).[5]

## 4. CASE STUDIES

In this section, we use two case studies to illustrate how to measure gender politics using computer vision: (a) 'woman card' and (b) Bernie Sanders followers jumping to Trump. All the profile images are classified using the neural network trained in Section III.

---

[3] http://opencv.org.
[4] The full list of label names together with the validation data set and the trained model, is available at the first author's website.
[5] The trained model has been deployed at the following image mining website: http://www.ifacetoday.com.

### 4.1 Woman Card

During his victory speech on April 26, 2016, Donald Trump accused Hillary Clinton of playing the 'woman card,' and said that she would be a failed candidate if she were a man. Clinton fired back during her victory speech in Philadelphia and said that "If fighting for women's health care and paid family leave and equal pay is playing the 'woman card,' then deal me in." The 'woman card' subsequently became the meme of the week and its effects are much debated. According to *CNN, New York Times, Washington Post* and *The Financial Times*, this exchange between the two presidential nominees signaled a heated general election clash over gender.[6]

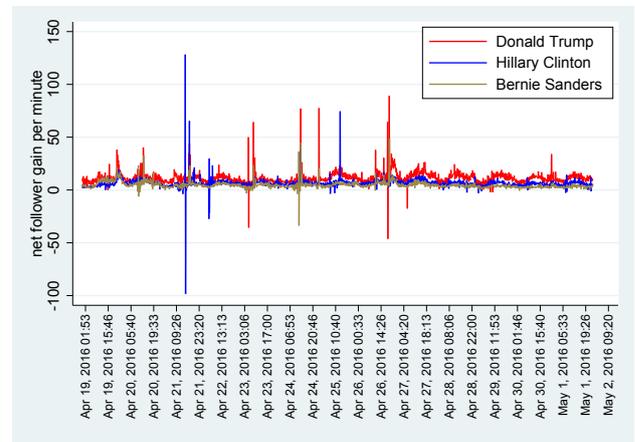

Figure 5: Net follower gain per minute for Trump, Clinton and Sanders between April 19 and May 2.

By leveraging the gender classifier and the detailed information on followers, we can easily measure the effects of the 'woman card' exchange on the gender composition of new followers and unfollowers for both Hillary Clinton and Donald Trump. Specifically, here we examine whether this exchange has made women more likely to follow Hillay Clin-ton and more likely to leave Trump.

---

[6] See, for example, http://www.nytimes.com/2016/04/29/us/politics/hillary-clinton-donald-trump-women.html.

Our dataset *US2016* contains the detailed IDs of Trump's and Clinton's followers. Specifically for this case study, we are able to use these IDs to identify all the new followers and the unfollowers of Donald Trump first between April 19 and April 26 and then between April 26 and May 1 (Figure 5). Similarly, we have information on Hillary Clinton's new followers and unfollowers first between April 20 and April 27 and then between April 27 and May 2. This enables us to examine in a definitive manner the gender composition of new followers and unfollowers one week before the 'woman card' exchange (April 26) and one week after. We report the summary statistics in Table 8.

Table 2: Mobility in the Candidates' Followers

|  | Hillary Clinton | | Donald Trump | |
| --- | --- | --- | --- | --- |
| 'Woman Card' | Before | After | Before | After |
| New Followers | 72,266 | 54,820 | 116,456 | 115,246 |
| Unfollowers | 9,572 | 8,393 | 18,376 | 18,292 |

Furthermore, as we have the follower information of other presidential candidates such as Bernie Sanders and Ted Cruz, we are able to identify the destinations of Trump and Clinton unfollowers. We report these statistics in Table 3 and Table 4.

Table 3: Mobility of Hillary Clinton's Unfollowers

| Destination | Bernie Sanders | Donald Trump | Ted Cruz* |
| --- | --- | --- | --- |
| Before | 14.55% | 11.95% | 2.19% |
| After | 12.47% | 11.03% | 2.62% |

*Ted Cruz has dropped out after the Indiana primary.

Table 4: Mobility of Donald Trump's Unfollowers

| Destination | Hillary Clinton | Bernie Sanders | Ted Cruz |
| --- | --- | --- | --- |
| Before | 6.04% | 4.87% | 4.55% |
| After | 5.94% | 4.54% | 3.70% |

### 4.1.1 New Followers

In Figure 6, we report on the gender composition of Clinton's new followers one week before the 'woman card' exchange and one week after. We observe a 1.6% increase in percentage of women followers. Our sample size is 14,504 in the first week and in the second 11,147.

In Figure 7, we report on the gender composition of Trump's new followers one week before the 'woman card' exchange and one week after. We observe a 0.6717% increase in percentage of women followers. Our sample size is 20,204 in the first week and in the second 21,187. While our main focus is the time-series variations for the candidates, it is interesting to note that across candidates, Clinton attracts more new female followers proportionally than Trump.

Using score test (Table 5), we are able to show that for Clinton the surge of female presence among her new followers is statistically significant. The same does not hold for Donald Trump.

### 4.1.2 Unfollowers

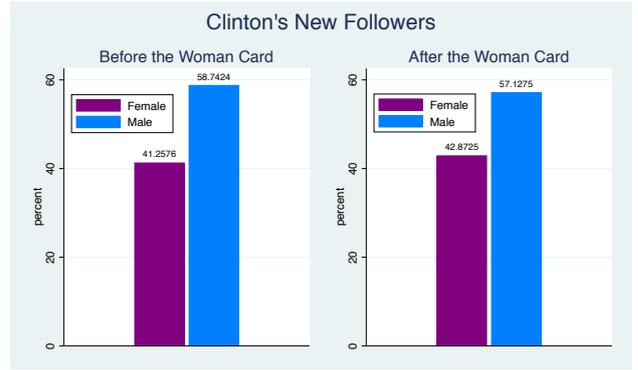

**Figure 6:** Gender Composition of Hillary Clinton's New Followers.

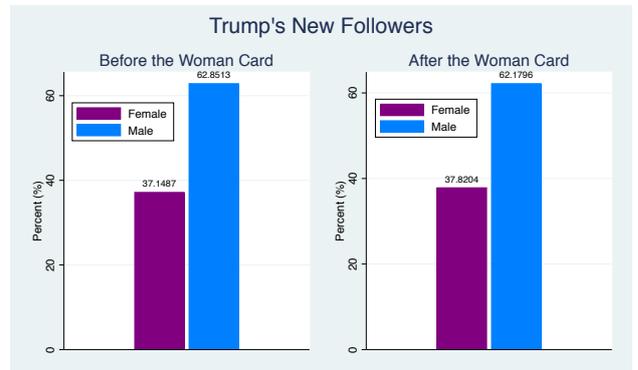

**Figure 7:** Gender Composition of Donald Trump's New Followers.

Table 5: New Followers' Gender Composition

| Null Hypothesis | Clinton | | Trump | |
| --- | --- | --- | --- | --- |
|  | z statistic | *p* value | z statistic | *p* value |
| $p_{before}=p_{after}$ | 2.597 | 0.0093 | 1.411 | 0.1582 |

In Figure 8, we report on the gender composition of Clinton's unfollowers one week before the 'woman card' exchange and one week after. We observe a 3.7728% decrease in the percentage of women unfollowers. Our sample size is 2039 in the first week and 1587 in the second.

In Figure 9, we report on the gender composition of Trump's unfollowers one week before the 'woman card' exchange and one week after. We observe a 0.2786% decrease in the percentage of women unfollowers. Our sample size is 3682 in the first week and 3036 in the second.

Using score test (Table 6), we show that for Clinton the decrease of female presence among her unfollowers is statistically significant at 95% confidence interval. While Donald Trump also observes a decrease in the percentage of female unfollowers, the decrease is not statistically significant.

Table 6: Unfollowers' Gender Composition

| Null Hypothesis | Clinton | | Trump | |
| --- | --- | --- | --- | --- |
|  | z statistic | *p* value | z statistic | *p* value |
| $p_{before}=p_{after}$ | -2.2581 | 0.0239 | -0.23178 | 0.8167 |

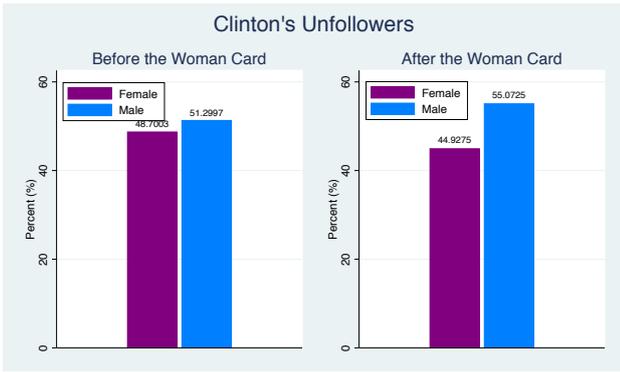

Figure 8: Gender Composition of Clinton's Unfollowers.

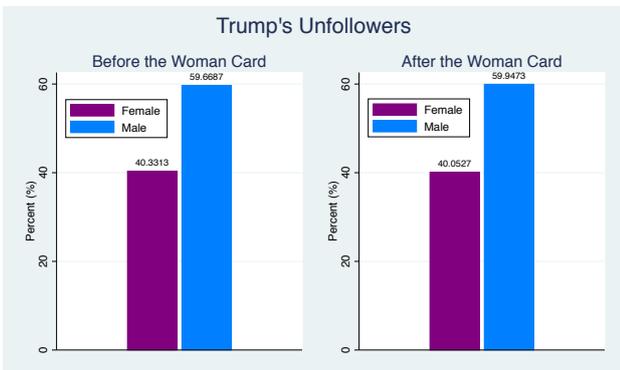

Figure 9: Gender Composition of Trump's Unfollowers.

## 4.2 Sanders Supporters Jumping Ship for Trump

Even as Hillary Clinton moves closer and closer to clinching the Democratic nomination, some Sanders supporters insist that they will not vote for her in the general election. This leaves a golden opportunity to Donald Trump, the Republican nominee who has been sharing similar campaign messages with Sanders on trade and campaign finance. Trump, on the other hand, also makes it clear that he will target Sanders supporters.[7]

This new dynamic quickly became hotly debated and the looming question is "Can Trump win over Sanders supporters?"[8] A number of recent polls, including ABC/Washington Post, CBS/NYT and YouGov, do suggest that some Sanders supporters could end up voting for Trump and that this is particularly so for his male supporters.[9] Here we investigate this new dynamic on Twitter.

Delving into the dataset *US2016*, we first examine the proportion of Sanders followers who simultaneously follow Trump or Clinton. Then we use our convolutional neural network to classify the followers' gender and study how men

---
[7]Newsweek, http://www.newsweek.com/can-trump-win-over-bernie-sanders-supporters-459218.
[8]Time, http://time.com/4420424/hillary-clinton-tim-kaine-running-mate-reaction.
[9]New York Times, http://www.nytimes.com/2016/05/25/upshot/explaining-hillary-clintons-lost-ground-in-the-polls.html.

Table 7: Number of Profile Images for Sanders Followers

|  | All Sanders | Sanders & Trump |
| --- | --- | --- |
| Number of Images | 40,088 | 34,921 |

in particular are responding to Trump's overture. We report the summary statistics in Table 8.

Table 8: Number of Followers

|  | March 24 | April 17 | May 10 |
| --- | --- | --- | --- |
| Bernie Sanders | 1,777,861 | 1,977,982 | 2,134,917 |
| Hillary Clinton | 5,755,618 | 5,905,124 | 6,176,731 |
| Donald Trump | 7,075,507 | 7,604,915 | 8,020,568 |

Using follower information, we first study among Sanders followers who are following Trump but not Clinton and who are following Clinton but not Trump. We think it is reasonable to assume that if Sanders drops out of the race, Sanders followers who are following Trump but not Clinton will support Trump, and that those who are following Clinton but not Trump will support Clinton. To match the candidates' millions of followers, we first sort their IDs and then use binary search. The entire matching process can be done within a few minutes.[10]

Next, we answer the question of whether Sanders followers are jumping ship for Trump. Specifically, we examine whether an increasing proportion of Sanders followers are now following Trump and, if so, whether this phenomenon is particularly significant for men.

### 4.2.1 Evolution of Sanders Followers

We first analyze the composition of Sanders followers between March and May. We divide Sanders followers into four groups: (1) only follow Trump and Sanders, (2) only follow Clinton and Sanders, (3) follow Trump, Clinton and Sanders, (4) only follow Sanders. While acknowledging that not all followers are supporters, we assume that followers in Group 1 are the most likely to switch to Trump and followers in Group 2 are the least likely. We report our results in Figures 8, 9, 10.

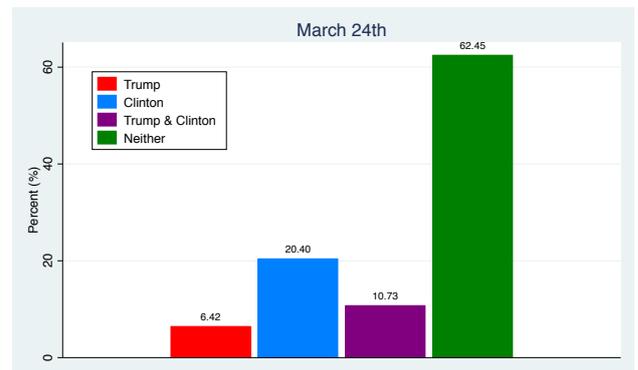

Figure 10: Composition of Sanders followers, March.

---
[10]Codes and data used in this paper are available on the first author's website.

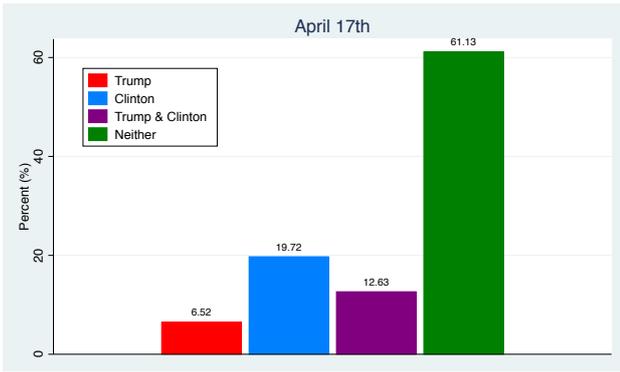

Figure 11: Composition of Sanders followers, April.

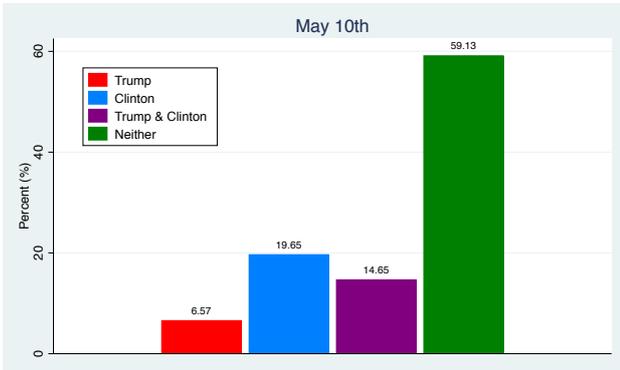

Figure 12: Composition of Sanders followers, May.

The results indicate a decrease of Clinton followers and an increase of Trump followers among Sanders followers between March and May. In Table 9, we use score test to show that the increase of Trump's presence and the drop of Clinton's presence are statistically significant.

Meanwhile, we also observe that individuals who follow only Sanders, marked by green, make up a smaller share in May than in March and that the share of individuals who follow all the three candidates has increased. Using score test, we are also able to show that these changes are statistically significant.

Table 9: The Composition of Sanders Followers

| Null Hypothesis | Clinton & Sanders | | Trump & Sanders | |
|---|---|---|---|---|
| | z statistic | $p$ value | z statistic | $p$ value |
| $\Pr_{March}=\Pr_{May}$ | -18.47 | 0.00 | 5.99 | 0.00 |

### 4.2.2 Gender

Having analyzed the composition of Sanders followers, we now focus specifically on Group 1, i.e. Sanders followers who also follow Trump but not Hillary Clinton. Our goal is to measure whether men are more likely to jump ship for Trump than women. Our investigation is motivated by poll findings that show white male supporters of Bernie Sanders are the most likely to switch to Trump.[11]

---

[11]Washington Post, https://www.washingtonpost.com/news/the-fix/wp/2016/05/24/how-likely-are-bernie-sanders-supporters-to-actually-vote-for-donald-trump-here-are-some-clues.

We use data collected on May 10th, when Sanders has 2,134,917 followers, of which 140,185 simultaneously follow Trump but not Clinton. Using the neural network reported in Section III, we classify the gender of these followers. We report the results in Figure 13.

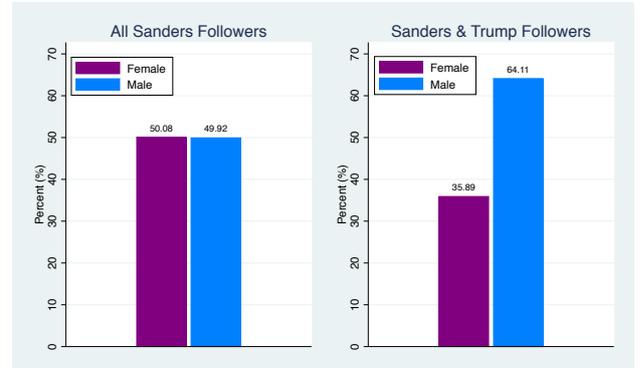

Figure 13: Sanders followers who are likely to jump ship for Trump are disproportionately male.

We find that of all Sanders followers 49.92% are male, but for those who follow both Sanders and Trump (and not Clinton) the percentage is as high as 64.11%. Using score test (Table 10), we show that among the Sanders followers, those who simultaneously follow Trump but not Clinton are more likely to be male than an average Sanders follower. Our results are consistent with the poll results and lend further support to previous studies that demonstrate the gender effect.

Table 10: Score Test on Gender Composition

| Null Hopythesis | Men | |
|---|---|---|
| | z statistic | $p$ value |
| $\Pr_{Sanders}=\Pr_{Sanders\ \&\ Trump}$ | 39.10 | 0.00 |

## 5. LIMITATIONS AND FUTURE RESEARCH

First, our work is built on the assumption that Twitter users, campaign followers in particular, are representative of the demographics of the U.S. population. This assumption may not exactly hold as various demographic dimensions such as gender, race and geography are skewed in Twitter [18]. Second, in our analysis we have deliberately removed empty profile images, as they are not informative with regards to gender. Posting an empty profile image might be correlated to one's following behavior. So both cases could potentially produce selection bias and affect our estimation[8]. Nonetheless, we believe the direction of our estimates will remain consistent, especially if calibrated by reliable polls.

## 6. CONCLUSIONS

Gender has been playing an important role in the U.S. presidential elections. Recent advances in computer vision,

on the other hand, have made gender classification increasingly accurate. In this paper we introduced computer vision to the study of gender politics on the web.

We first collected all the profile images of the candidates' Twitter followers. Then we trained a highly accurate convolutional neural network using images that contain gender labels. Lastly, we classified all the follower and unfollower images. Through two case studies, we demonstrate how gender is informing the 2016 U.S. presidential election.

Our framework of analysis, which marries gender politics with computer vision, can be readily generalized to study other cases and other elections, such as the upcoming French presidential election in 2017. Our study has focused exclusively on images and we have demonstrated its effectiveness. We suggest, however, incorporating text-based analysis (e.g. of user name and tweets) [2] could be beneficial as well.

## 7. ACKNOWLEDGEMENT

We acknowledge support from the University of Rochester, New York State through the Goergen Institute for Data Science, and our corporate sponsors Xerox and Yahoo.